\documentclass[twocolumn]{jpsj2} %% two-column layout
\title{Microscopic Mechanism and Pairing 
Symmetry of Superconductivity in the
Noncentrosymmetric Heavy Fermion Systems 
CeRhSI$_3$ and CeIrSi$_3$}

\author{Yasuhiro \textsc{Tada}$^{1}$\thanks{E-mail address: tada@tp.ap.eng.osaka-u.ac.jp}, Norio \textsc{Kawakami}$^{1,2}$ and 
Satoshi \textsc{Fujimoto}$^{2}$ }

\inst{$^{1}$Department of Applied Physics, Osaka University, Suita, Osaka 565-0871, Japan \\
$^{2}$Department of Physics, Kyoto University, Kyoto 606-8502, Japan}

\abst{We study the pairing symmetry of the noncentrosymmetric 
heavy fermion superconductors CeRhSi$_3$ and CeIrSi$_3$ under pressures,
which are both antiferromagnets at ambient pressure.
We solve the Eliashberg equation by means 
of the random phase approximation and find that the mixed state of 
 extended $s$-wave and $p$-wave 
rather than the $d+f$ wave state could be realized 
by enhanced antiferromagnetic spin fluctuations. 
It is  elucidated that the gap function has line nodes on the Fermi surface
and the resulting density of state in the superconducting
state shows a similar character to that of usual $d$-wave superconductors, 
resulting in the NMR relaxation rate 
$1/(T_1T)$ that exhibits no coherence peak and behaves like
$1/(T_1T)\propto T^2$ at low temperatures. 
}

\kword{superconductivity, heavy fermion, without inversion
 symmetry}

\begin{document}
\maketitle

%%%%%%%%%%%%%%%%%%%%%%%%%%%%%%%%%%%%%%%%%%%%%
%introduction
%%%%%%%%%%%%%%%%%%%%%%%%%%%%%%%%%%%%%%%%%%%%%%
\section{Introduction}
 Recent discoveries of heavy fermion superconductors without 
inversion symmetry have attracted much interest. 
CePt$_3$Si\cite{pap:Bauer} 
was first identified and accompanied by 
the subsequent discoveries of CeRhSi$_3$\cite{pap:Kimura}, 
CeIrSi$_3$\cite{pap:Sugitani}, 
UIr\cite{pap:Akazawa} and CeCoGe$_3$\cite{pap:Settai}. 
Besides these heavy fermion systems, 
non-heavy fermion materials such as Li$_2$Pd$_3$B and Li$_2$Pt$_3$B 
were also found\cite{pap:Togano}. 
In all these materials, there are nonzero 
potential gradient 
$\nabla V$ averaged 
in the unit cell due to lack of inversion symmetry, which 
results in the anisotropic spin-orbit interaction 
expressed as $e\hbar/4m^2c^2(\mbox{\boldmath $k$}\times
\nabla V)\cdot \mbox{\boldmath $\sigma$}$ 
$(\equiv \alpha \mbox{\boldmath ${\cal L}$}_0\cdot \mbox{\boldmath $\sigma$})$,
where
$\mbox{\boldmath $k$}$ is the momentum of a particle and 
$\mbox{\boldmath $\sigma$}$ is Pauli matrices.
The anisotropic spin-orbit interaction
$\mbox{\boldmath ${\cal L}$}_0\cdot \mbox{\boldmath $\sigma$}$,
 whose general form can be determined by a group theoretical argument
\cite{pap:Samokhin}, leads to many interesting phenomena
\cite{pap:Edelstein,pap:Edelstein2,pap:Gorkov,pap:Yip,
pap:Sigrist,pap:Frigeri,pap:Kaur,pap:Fujimoto,pap:Fujimoto2,
pap:Fujimoto3,pap:Yanase1}. 
One of the outstanding properties is the parity mixing 
in superconducting states\cite{pap:Edelstein,pap:Gorkov,pap:Frigeri,pap:Sigrist,pap:Fujimoto3,
pap:Yanase1},
{\it i.e.} the admixture of the 
spin-singlet and triplet states, which are both well 
defined in superconductors with inversion symmetry. 
The pairing symmetry in CePt$_3$Si has been 
studied both theoretically
\cite{pap:Sigrist,pap:Frigeri,pap:Fujimoto,pap:Fujimoto3,
pap:Yanase1} and experimentally
\cite{pap:Yogi,pap:Yogi2,pap:Izawa,pap:Takeuchi,
pap:Bonalde} and it is believed that the $s+p$ wave
superconducting state is realized. 
Frigeri $et$ $al.$\cite{pap:Frigeri} pointed out
that the spin-orbit interaction could determine the direction
of the $\mbox{\boldmath $d$}$-vector as 
$\mbox{\boldmath $d$}\propto \mbox{\boldmath ${\cal L}$}_0
$ for which the highest
transition temperature was obtained.
A microscopic calculation with the detailed structure of
the Fermi surface was done\cite{pap:Yanase1} and 
it was concluded that $s+p$ wave state is the most 
probable state.

 Among this new kind of compounds, CeRhSi$_3$ and CeIrSi$_3$ 
have many similarities due to the same crystal 
structure: qualitatively similar pressure-temperature 
phase diagrams were indeed obtained from resistivity measurements
\cite{pap:Kimura,pap:Sugitani}. 
They are both in antiferromagnetic (AF) ordered states at low pressures,
which are driven to the superconducting states beyond 
 the critical pressures where the Neel temperature rapidly decreases. 
The Neel temperature at ambient pressure is 
$T_N=1.6$K ($5.0$K) for CeRhSi$_3$ (CeIrSi$_3$) and 
the superconductivity appears in wide pressure ranges with 
approximate maximal transition temperature
$1.1$K at $2.6$GPa ($1.6$K at $2.5$GPa), respectively.
Moreover, the NMR measurements of the 
relaxation rate $1/T_1$  for CeIrSi$_3$\cite{com:Mukuda} suggest 
the existence of the AF spin fluctuations. Also, the
neutron scattering experiments for CeRhSi$_3$\cite{pap:Aso} 
identified the AF ordering vectors 
as $\mbox{\boldmath $Q$}=\left( \pm 0.215,0,0.5\right)$ and 
it is concluded that the character of the AF is SDW-like. 
Besides these experiments, the recent band calculations
 elucidated that the above two compounds have very similar 
Fermi surfaces\cite{com:Harima,pap:Terashima}(FS). 
These similarities deduced both experimentally 
and theoretically  motivate us to discuss 
the superconductivities of these two 
compounds in the same theoretical framework.

 In this paper, we study the noncentrosymmetric superconductors
 CeRhSi$_3$ and CeIrSi$_3$ with particular emphasis on
 the influence of AF fluctuations
to identify the pairing symmetry
realized in these systems.  We also examine the
properties in the superconducting state, the
density of states and the NMR relaxation rate which 
characterize the nodal structure
of a gap function on the FS.  

This paper is organized as follows. In the next section we 
introduce the model and briefly mention basic properties in the
non-interacting case. Then in \S 3, we examine possible types of
pairing symmetry by means of the random phase approximation.
In \S 4, the characteristic properties in the
superconducting state are discussed, and a brief summary 
is given in \S 5. 

%%%%%%%%%%%%%%%%%%%%%%%%%%%%%%%%%%%%%%%%%
%model@@@
%%%%%%%%%%%%%%%%%%%%%%%%%%%%%%%%%%%%%%%%%%%
\section{Model}
\begin{figure}[tb]
\begin{center}
\includegraphics[width=0.5\linewidth]{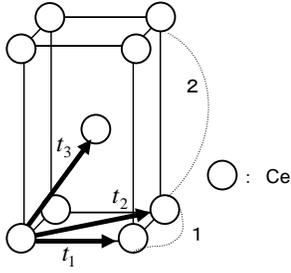}
\end{center}
\caption{Unit cell of CeRh(Ir)Si$_3$: only
Ce sites (open circles) are shown for clarity, 
which form a body centered tetragonal lattice.
$t_1$, $t_2$ and $t_3$ are the nearest, the second nearest
and the third nearest-neighbor hopping integrals, respectively.
The ratio of the lattice constants is $a:b:c=1:1:2$.}
\label{fig:BCT}
\end{figure}

In CeRhSi$_3$ and CeIrSi$_3$, heavy 4$f$-electrons 
around the Fermi level
play important roles for low energy phenomena, where
4$f$ electrons come from Ce ions forming
 a body-centered tetragonal(BCT) lattice\cite{pap:Kimura}
as shown in Fig. \ref{fig:BCT}. 
%%%%figBCT%%%%%%%%%%%%%
%%%%%%%%%%%%%%%%%%%%%%
For our analysis of the superconductivity, 
we start with the situation that heavy fermions 
have already been formed through hybridizations with 
conduction electrons and are described 
by an effective Hamiltonian. 
Although these materials have two kinds of the Fermi surfaces
\cite{com:Harima,pap:Terashima,pap:Yamagami} apart from the splitting 
by the spin-orbit interaction, we focus on
one of them that has a large weight of the total density of states. 
This enables us to simply describe the electrons in the
materials by the following single band model, 

%%%%%%%%%%%%%%%%%%%%%%%%%%%%%%%%
\begin{eqnarray}
 H&=&\sum_{k}\varepsilon_{k}
 c^{\dagger}_{k}c_{k}
 +U\sum_{i}n_{i\uparrow}n_{i\downarrow} \nonumber \\
 && \quad +\alpha \sum_{k}c^{\dagger}
 _{k}\mbox{\boldmath ${\cal L}$}_0
 \left(\mbox{\boldmath $k$}\right)
 \cdot \mbox{\boldmath $\sigma$} c_{k},\\
 \label{hamiltonian}
 %%%%%%%%%%%%%%%%%%%%%%%%%%%%%
 \mbox{\boldmath ${\cal L}$}_0\left(\mbox{\boldmath $k$}\right)
 &=&\left(\sin k_y,-\sin k_x,0\right),\\
 \varepsilon_{k}&=&-2t_1(\cos k_x+\cos k_y)+4t_2
 \cos k_x\cos k_y \nonumber \\
 &&-8t_3\cos (k_x/2)\cos (k_y/2)\cos k_z-\mu ,
\end{eqnarray}
%%%%%%%%%%%%%%%%%%%%%%%%%%%%%%%%%%%%%%%%%
where $c^{(\dagger)}_k=\left(c_{k\uparrow},
c_{k\downarrow}\right)^{t(\dagger)}$ are the annihilation (creation)
operators of the Kramers doublet. The third term 
of (\ref{hamiltonian})  is the Rashba-type anisotropic spin-orbit 
interaction due to the lack of inversion symmetry,
where its coupling constant $\alpha$
 is estimated to be less than $0.1\varepsilon_{F}$
($\varepsilon_{F}$ is the Fermi energy)
according to the band calculation\cite{com:Harima}.
The bare Green's function in the normal state is 
%%%%%%%%%%%%%%%%%%%%%%
\begin{eqnarray*}
\hat{G}^0(k)&=&\sum_{\tau=\pm}\frac{\sigma_0+\left( 
\mbox{\boldmath ${\cal L}$}_0(\mbox{\boldmath $k$})/
\| \mbox{\boldmath ${\cal L}$}_0(\mbox{\boldmath $k$})\| \right)\cdot
	\mbox{\boldmath ${\sigma}$}}{2}G_{\tau}^0(k),\\
G_{\tau}^0(k)&=&\frac{1}{i\omega_n-\xi_{k\tau}},\\
\xi_{k\tau}&=&\varepsilon_k
 +\tau \alpha \| \mbox{\boldmath ${\cal L}$}_0(\mbox{\boldmath $k$})\|,\\
\| \mbox{\boldmath ${\cal L}$}_0(\mbox{\boldmath $k$})\|&=&
 \sqrt{{\cal L}_{0x}^2+{\cal L}_{0y}^2+{\cal L}_{0z}^2},
\end{eqnarray*}
%%%%%%%%%%%%%%%%%%%%%%%%
where $k=(i\omega_n,
\mbox{\boldmath $k$})$.
Note that the Green's functions have nonzero off-diagonal
elements because of the spin-orbit interaction.

\begin{figure}[tb]
\begin{center}
\includegraphics[width=0.7\linewidth]{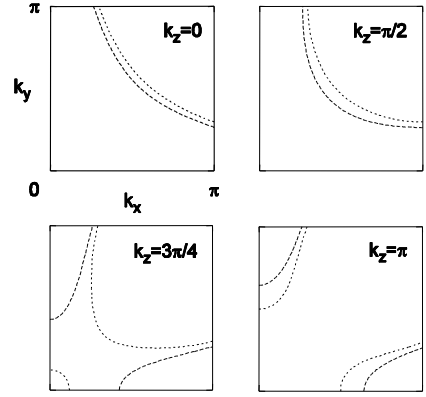}
\end{center}
\caption{Fermi surfaces in our model.
The cross sections at $k_z=0$, $k_z=\pi/2$, $k_z=3\pi/4$
and $k_z=\pi$ are shown.}
\label{fig:FS}
\end{figure}

\begin{figure}[tb]
\begin{center}
\includegraphics[width=0.9\linewidth]{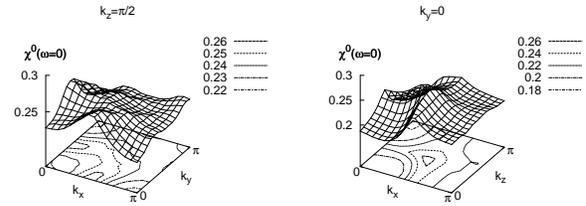}
\end{center}
\caption{Bare susceptibility $\chi^0(q)$ at $\omega =0$
for $\alpha =0$. The result is obtained 
for $T=0.04$. The left (right) panel is $\chi^0(q)$ on the $xy$
plane at $k_z=\pi/2$ ($xz$ plane at $k_y=0$)}
\label{fig:chi_0}
\end{figure}

\begin{figure}[tb]
\begin{center}
\includegraphics[width=0.9\linewidth]{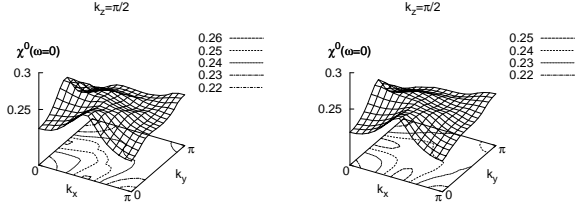}
\end{center}
\caption{Bare susceptibility $\chi^0(q)$
on the $xy$ plane at $k_z=\pi/2$ and $\omega =0$
for $\alpha \neq0$ ($T=0.04$). 
The left (right) panel is for $\alpha=0.1$ (0.2).}
\label{fig:chi_12}
\end{figure}

With a natural assumption 
deduced from the above-mentioned experimental and theoretical results 
that the AF in these materials are driven by the 
nesting of the Fermi surfaces, 
we choose the parameters in the above Hamiltonian 
$t_1,t_2,t_3$ and filling $n$ so that our model should be
consistent with the band calculation and the neutron 
scattering experiment; $(t_1,t_2,t_3,n)=(1.0,0.475,0.3,1.055)$.
Here, we define $t_1$ as the energy unit. 
For these fixed parameters, the split Fermi surfaces 
with spin-orbit splitting $\sim 2\alpha$ are 
shown in Fig. \ref{fig:FS}.
%%added
Our tight-binding model successfully reproduces
the characteristic features of the Fermi surfaces obtained by
the first principle band calculations.
\cite{com:Harima,pap:Terashima,pap:Yamagami}
%%%
The matrix elements of the momentum-dependent
susceptibility $\hat{\chi}^0(q)$ at $U=0$ are
expressed as, 
%%%%%%%%%%%%%%%%%%%%%%%%%%%
\begin{eqnarray}
\chi^0_{s_1s_2s_3s_4}(q)=-\frac{T}{N}\sum_{k}
G^0_{s_2s_1}(q+k)G^0_{s_4s_3}(k).
\end{eqnarray}
%%%%%%%%%%%%%%%%%%%%%%%%%%
We show the bare susceptibility 
$\chi^0(q) \equiv \hat{\chi}^0_{\uparrow \uparrow \uparrow \uparrow}(q)$ for 
several choices of $\alpha$ at $\omega=0$
in Figs. \ref{fig:chi_0} and \ref{fig:chi_12}.
At $\alpha=0$, It has peaks 
around $\mbox{\boldmath $Q$}_{1\pm}\sim (\pm \pi/2,0,\pi/2)$,
$\mbox{\boldmath $Q$}_{2\pm}\sim (0,\pm \pi/2,\pi/2)$ as 
shown in Fig. \ref{fig:chi_0}.
When the spin-orbit coupling $\alpha$ is turned on, the 
anisotropic spin-orbit splitting emerges, which 
 suppresses the nesting of the FS and thus
slightly affects the behavior of $\chi^0$:
the peak structure is a little bit smeared,
as seen from Fig. \ref{fig:chi_12}.
%%%
The peak structure in $\hat{\chi}^0(q)$ is 
qualitatively in agreement with 
the neutron scattering experiment\cite{pap:Aso},
although its profile in momentum space is
not so sharp in our model.

%%%%figFS%%%%%%%%%%%%%
%%%%%%%%%%%%%%%%%%%%%%

%%%%fig:chi0%%%%%%%%%
%%%%%%%%%%%%%%%%%%%

%%%%fig:chi12%%%%%%%%%
%%%%%%%%%%%%%%%%%%%

%%%%%%%%%%%%%%%%%%%%%%%%%%%%%%%%%%%%%%%%%%%
%paring symmetry@@@
%%%%%%%%%%%%%%%%%%%%%%%%%%%%%%%%%%%%%%%%%%%%
\section{Pairing symmetry}
 In this section, we study the pairing symmetry by solving the
Eliashberg equation which is expressed as
%%%%%%%%%%%%%%%%%
\begin{eqnarray}
 \lambda 
 \Delta_{s_1s_2}(k)&=&-\frac{T}{N}\sum_{k^{\prime}s_3s_4\sigma_1\sigma_2}
  V_{s_1s_2s_3s_4}(k,k^{\prime}) \nonumber \\
 &&\quad \times G^0_{\sigma_1s_3}(k^{\prime})G^0_{\sigma_2s_4}(-k^{\prime})
 \Delta_{\sigma_1\sigma_2}(k^{\prime}).
\label{eq:Eliashberg}
\end{eqnarray}
%%%%%%%%%%%%%%%%%
where $\lambda$ is the eigenvalue.
We use the random phase approximation (RPA), which  
incorporates large AF fluctuations at the pressures 
near the AF critical point, 
to evaluate the effective pairing interaction $V$. 
In the following discussions, we neglect the normal 
self energy, which may have
 little influence on determining the pairing symmetry
\cite{pap:Yanase2}.
%%%%%%%%%%%%%%%%%%%%%%%%%%%%%%%%%%%%%%
The effective pairing interaction $V$ 
evaluated within RPA consists of three parts, 
%%%%%%%%%%%%%%%%%%%%%%%%%%%%%%%
\begin{eqnarray}
V_{s_1s_2s_3s_4}(k,k^{\prime})&=&
U\delta_{s_1s_3}\delta_{s_2s_4}\delta_{s_1\bar{s_2}} \nonumber \\
&&+V^{{\rm bub}}_{s_1s_2s_3s_4}(k,k^{\prime})
 +V^{{\rm lad}}_{s_1s_2s_3s_4}(k,k^{\prime}), \nonumber \\
\end{eqnarray}
%%%%%%%%%%%%%%%%%%%%%%%%%%%%%%%%%
where $U$ is the bare Hubbard repulsion and 
$V^{{\rm bub}}$, $V^{{\rm lad}}$ are calculated by 
collecting bubble and ladder diagrams, respectively.
The bubble terms are
\begin{eqnarray*}
V^{{\rm bub}}_{ssss}(k,k^{\prime})
&=&v^{{\rm bub}}_{ssss}(k-k^{\prime})-v^{{\rm bub}}_{ssss}(k+k^{\prime}),\\
V^{{\rm bub}}_{s\bar{s}s\bar{s}}(k,k^{\prime})
&=&v^{{\rm bub}}_{s\bar{s}s\bar{s}}(k-k^{\prime}),
\end{eqnarray*}
where
\begin{eqnarray*}
&&v^{{\rm bub}}_{ssss}(q)=
-U^2\chi^0_{\bar{s}\bar{s}\bar{s}\bar{s}}(q)/D_{{\rm bub}}(q),\\
&&v^{{\rm bub}}_{s\bar{s}s\bar{s}}(q)=
U^2\bigl( -\chi^0_{\bar{s}ss\bar{s}}(q)
-U\chi^0_{\bar{s}ss\bar{s}}(q)
\chi^0_{s\bar{s}\bar{s}s}(q)\\
&&\qquad \qquad+U\chi^0_{\bar{s}\bar{s}\bar{s}\bar{s}}(q)
\chi^0_{ssss}(q)\bigr)/D_{{\rm bub}}(q),\\
&&D_{{\rm bub}}(q)=\left(1+U\chi^0
_{\downarrow\uparrow\uparrow\downarrow}(q)\right)
\left(1+U\chi^0_{\uparrow\downarrow\downarrow\uparrow}(q)\right)\\
&&\qquad \qquad-U^2\chi^0_{\uparrow\uparrow\uparrow\uparrow}(q)
\chi^0_{\downarrow\downarrow\downarrow\downarrow}(q)
\end{eqnarray*}
and $V^{{\rm bub}}_{s_1s_2s_3s_4}$ with other spin indices
are zero.
The ladder terms are
%%%%%%%%%%%%%%%%%%%%%%%%5
\begin{eqnarray*}
V^{{\rm lad}}_{ss\bar{s}\bar{s}}(k,k^{\prime})
&=&v^{{\rm lad}}_{ss\bar{s}\bar{s}}(k-k^{\prime})
-v^{{\rm lad}}_{ss\bar{s}\bar{s}}(k+k^{\prime}),\\
V^{{\rm lad}}_{s\bar{s}s\bar{s}}(k,k^{\prime})
&=&v^{{\rm lad}}_{s\bar{s}s\bar{s}}(k-k^{\prime}),
\end{eqnarray*}
%%%%%%%%%%%%%%%%%%%%%%%%%
where
%%%%%%%%%%%%%%%%%%%%%%%%%%%%
\begin{eqnarray*}
&&v^{{\rm lad}}_{ss\bar{s}\bar{s}}(q)=
U^2\chi^0_{s\bar{s}s\bar{s}}(q)/D_{{\rm lad}}(q),\\
&&v^{{\rm lad}}_{s\bar{s}s\bar{s}}(q)=
U^2\bigl(\chi^0_{ss\bar{s}\bar{s}}(q)
-U\chi^0_{\bar{s}\bar{s}ss}(q)\chi^0_{ss\bar{s}\bar{s}}(q)\\
&&\qquad \qquad 
+U\chi^0_{s\bar{s}s\bar{s}}(q)\chi^0_{\bar{s}s\bar{s}s}
(q)\bigr)/D_{{\rm lad}}(q),\\
&&D_{{\rm lad}}(q)=\left(1-U\chi^0
_{\downarrow\downarrow\uparrow\uparrow}(q)\right)
\left(1-U\chi^0_{\uparrow\uparrow\downarrow\downarrow}(q)\right)\\
&&\qquad \qquad-U^2\chi^0_{\uparrow\downarrow\uparrow\downarrow}(q)
\chi^0_{\downarrow\uparrow\downarrow\uparrow}(q)
\end{eqnarray*}
%%%%%%%%%%%%%%%%%%%%%%%%%%%%%%%
and others are zero. 
Within RPA, only the $V^{{\rm lad}}_{ss\bar{s}\bar{s}}$ 
terms do not conserve the spins of two particles before and after 
scattering, and the parity mixing is driven only by two Green's 
functions which 
connect the 4-point vertex part with the gap function in 
eq.(\ref{eq:Eliashberg}).
Thus, the parity mixing effect may not be strongly enhanced by $U$.

Generally, the gap function $\Delta_{s_1s_2}\left( k\right)$ is 
expressed as, 
%%%%%%%%%%%%%%%%%%%%%%%%%%%%%%%%%%%%
\begin{eqnarray}
 \Delta \left( k\right)=\bigl( \Delta_s(i\omega_n)
 d_0( \mbox{\boldmath $k$})
 \sigma_0+\Delta_t(i\omega_n)
 \mbox{\boldmath $d$}( \mbox{\boldmath $k$})
 \cdot \mbox{\boldmath $\sigma$}
 \bigr)i\sigma_2,
\end{eqnarray}
%%%%%%%%%%%%%%%%%%%%%%%%%%%%%%%%%%%
where $\Delta_sd_0$ and 
$\Delta_t\mbox{\boldmath $d$}$ 
are the order parameters for singlet and triplet states, which 
can have non-zero values simultaneously because of 
the Rashba spin-orbit interaction. 
We denote $\left( d_{\mu}\right)_{\mu =0\sim3}
=\left( d_0,\mbox{\boldmath $d$}\right)$ hereafter.
We can determine, by solving the above eigenvalue 
equation (\ref{eq:Eliashberg}), 
the symmetry of the gap function and 
the transition temperature $T_c$ at which 
the maximum eigenvalue $\lambda_{{\rm max}}$ reaches unity.

 The pairing symmetries of the singlet states 
$\{ d_0^{\Gamma}(\mbox{\boldmath $k$})\}_{\Gamma}$ 
for five irreducible representations of 
C$_{4v}$ are listed in Table \ref{t1}.
%%%%%%%%%%%%%%%%table 1%%%%%%%%%%
\begin{table}[htbp]
\begin{tabular}{ll}
\hline
\hline
Irreducible & $\quad$  \\
representation & Basis function \\ \hline
A$_1$(extended $s$) 
& $d_0^{{\rm A}_1}(\mbox{\boldmath $k$})=\cos 2k_z$ \\ 
A$_2$($g_{xy(x^2-y^2)}$) 
& $d_0^{{\rm A}_2}(\mbox{\boldmath $k$})=
\sin 2k_x\sin 2k_y\left( \cos 2k_x-\cos 2k_y\right)$ \\
B$_1$($d_{x^2-y^2}$) 
& $d_0^{{\rm B}_1}(\mbox{\boldmath $k$})=\left( \cos 2k_x-\cos 2k_y\right)$ \\
B$_2$($d_{xy}$) 
& $d_0^{{\rm B}_2}(\mbox{\boldmath $k$})=\sin 2k_x\sin 2k_y$ \\
E  ($d_{xz}$) 
& $d_0^{{\rm E}}(\mbox{\boldmath $k$})=\sin k_x\sin 2k_z$ \\
 & \\
every representation & $\mbox{\boldmath $d$}^{\Gamma}(\mbox{\boldmath $k$})
=d_0^{\Gamma}(\mbox{\boldmath $k$})
\mbox{\boldmath ${\cal L}$}_0(\mbox{\boldmath $k$})$ \\
\hline
\end{tabular}
\caption{The irreducible representations of C$_{4v}$ and 
 the basis functions. 
 }
\label{t1}
\end{table}
%%%%%%%%%%%%%%%%%%%%%%%%%%%%%%%%%%%%%%%%%%%
Here, we choose $\{\mbox{\boldmath $d$}
^{\Gamma}(\mbox{\boldmath $k$})\}_{\Gamma}$
as $\mbox{\boldmath $d$}^{\Gamma}(\mbox{\boldmath $k$})
=d_0^{\Gamma}(\mbox{\boldmath $k$})
\mbox{\boldmath ${\cal L}$}_0(\mbox{\boldmath $k$})$ for each representation 
which is considered to be most stable in the superconductors with 
$\Delta \ll \alpha$\cite{pap:Frigeri,pap:Fujimoto}.
The harmonic wave functions in Table \ref{t1} give the largest contributions 
among the functions which belong to a given symmetry, 
mainly because of 
the factor $1/2$ in the propagating vectors 
$\mbox{\boldmath $Q$}_{1,2\pm}$.

\begin{figure}[tb]
\begin{center}
\includegraphics[width=0.7\linewidth]{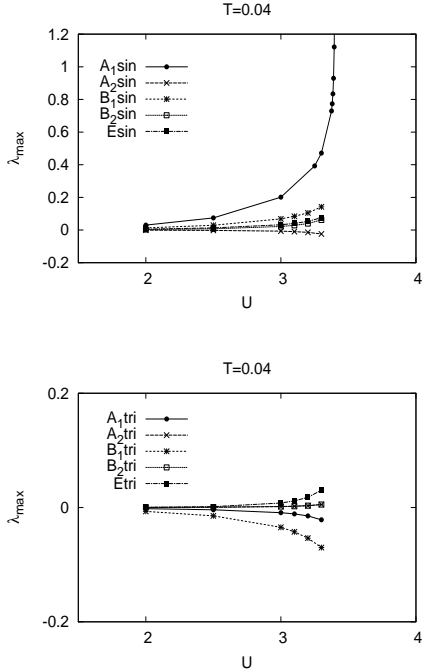}%{lambda_0_2.eps}
\end{center}
\caption{Maximum eigenvalues of the Eliashberg
equation for five irreducible representations 
as a function of $U$ with
$\alpha=0$ and $T=0.04$.
The upper (lower) panel is $\lambda_{\rm max}^{\Gamma}$ for the
singlet (triplet) gap function.
}
\label{fig:lambda_0}
\end{figure}

We solve the Eliashberg equation for all symmetries and 
trace each maximum eigenvalue $\{ \lambda^{\Gamma}\}_{\Gamma}$ 
with increasing $U$
at fixed temperature $T=0.04$. 
In the calculation, the first Brillouin zone is 
divided into $16\times16\times16$ meshes and the 
number of the Matsubara frequencies used is 512.
We have checked that the following results are 
qualitatively unchanged for 512 Matsubara frequencies and
$(32)^3$ 
$\mbox{\boldmath $k$}$-meshes.
In Fig. \ref{fig:lambda_0},
$\lambda (U,T=0.04)$ at $\alpha =0$ are shown
for five irreducible representations. In the case of $\alpha =0$,
the Eliashberg equation (\ref{eq:Eliashberg}) is 
separated into singlet and triplet parts and 
solved independently.
For singlet superconductivity,
we can see that, 
among five symmetries, only
$\lambda^{{\rm A}_1}_{\rm sin}$ for $d_0^{{\rm A}_1}=\cos 2k_z$
(extended $s$-wave)
can reach unity. 
Other $\lambda_{\rm sin}$ are much smaller 
than $\lambda^{{\rm A}_1}_{\rm sin}$, and
we cannot see any significant
difference among them in the present calculation.
Regarding the triplet part, all $\lambda_{\rm tri}$
are small and none of them can reach unity.

%%%%figlambda_0%%%%%%%%%%%%%
%%%%%%%%%%%%%%%%%%%%%%

We remark that, for A$_1$ representation, $\lambda^{{\rm A}_1}_{{\rm tri}}$
for $\mbox{\boldmath $d$}^{{\rm A}_1}(\mbox{\boldmath $k$})
=d_0^{{\rm A}_1}(\mbox{\boldmath $k$})
\mbox{\boldmath ${\cal L}$}_0(\mbox{\boldmath $k$})$
is very small and negative, which means that the triplet
($p$-wave) channel
in the effective interaction $V$ is weakly repulsive.

\begin{figure}[tb]
\begin{center}
\includegraphics[width=0.7\linewidth]{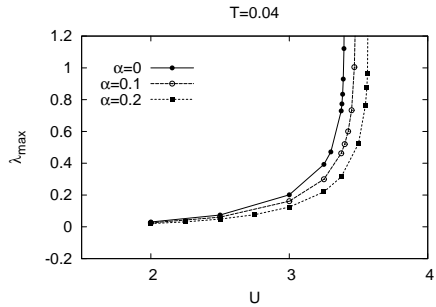}
\end{center}
\caption{Maximum eigenvalues $\lambda^{{\rm A}_1}$ for
$\alpha=0,0.1,0.2$ at $T=0.04$.
For $\alpha=0$, $\lambda$ is shown only for the singlet 
gap function, while for $\alpha \neq 0$, $\lambda$ is for
the singlet-triplet mixing state.}
\label{fig:lambda_12}
\end{figure}

Let us now inspect 
the effect of 
anisotropic spin-orbit interaction for A$_1$ symmetry, 
by computing $\lambda (U,T=0.04)$ at $\alpha=0,0.1,0.2$.
The results are shown in Fig. \ref{fig:lambda_12}. 
%%%%figlambda_12%%%%%%%%%%%%%
%%%%%%%%%%%%%%%%%%%%%%
Critical values of $U$, which correspond to 
the critical pressure for 
the AF transition, increase a little with $\alpha$, reflecting
 the change of FS that has a tendency to suppress $\hat{\chi}^0$.
Note that, within RPA calculations, the effective 
Coulomb interaction
 $U_{{\rm eff}}=U/{\rm min}(1-U\hat{\chi}^0(Q))$
is  a relevant parameter controlling
 spin fluctuations even in the presence of $\alpha$.
We see that $\lambda^{{\rm A}_1}$ decreases with 
increasing $\alpha$ mainly 
due to the
suppressed $U_{{\rm eff}}$ at the same value of $U$
and also
due to the mixing with
the repulsive triplet channel. 
However, it remains larger than unity, suggesting the possibility
of the A$_1$-symmetric superconductivity.
The ratio of the amplitudes of the singlet gap function
and the triplet gap function 
for A$_1$ symmetry is very small 
$\Delta_t/\Delta_s\lesssim 0.01$ for $\alpha \neq 0$,
which means that
the properties of the superconductivity is
characterized dominantly by the singlet part.

Generally, the A$_1$ symmetric gap function has line nodes perpendicular 
to $c$-axis on
the Fermi surface, but the averaged value over the FS 
$\langle d_{\mu}^{{\rm A}_1}\rangle_{{\rm FS}}$ has 
a non-zero value. 
In our model with this gap function,
there exist line nodes at $k_z=\pm \pi/4,\pm 3\pi/4$ 
on the FS. 
Regarding the nodal structure, our
 A$_1$ gap function $d_{\mu}^{{\rm A}_1}$ 
is very similar to that for the usual 
$d$-wave superconductivity.

\begin{figure}[tb]
\begin{center}
\includegraphics[width=0.9\linewidth]{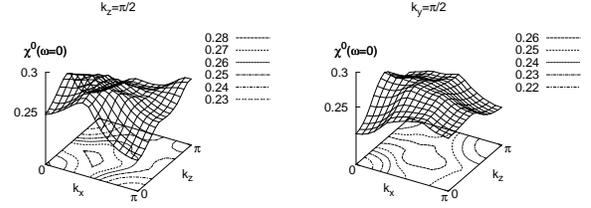}
\end{center}
\caption{Bare susceptibility 
$\chi^0(q)\equiv 
\chi^0_{\uparrow\uparrow\uparrow\uparrow}(q)$ 
 at $\omega =0$ and $T=0.04$
with $(t_1,t_2,t_3,n,\alpha) =(1.0,0.405,0.3,1.057,0.1)$.
 The left (right) panel is $\chi^0(q)$
on the $xy$
plane at $k_z=\pi/2$ (the $xz$ plane at $k_y=\pi/2$).}
\label{fig:chi_fs}
\end{figure}

\begin{figure}[tb]
\begin{center}
\includegraphics[width=0.7\linewidth]{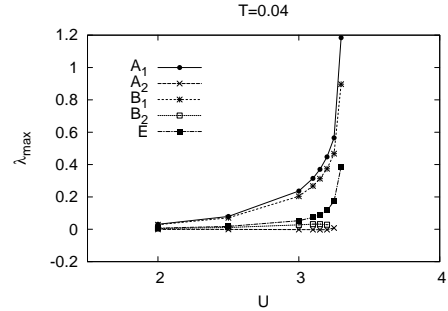}
\end{center}
\caption{Maximum eigenvalues $\lambda$ for
five irreducible representations with
$(t_1,t_2,t_3,n,\alpha) =(1.0,0.405,0.3,1.057,0.1)$
at $T=0.04$.}
\label{fig:lambda_fs}
\end{figure}

Because the neutron scattering 
experiment\cite{pap:Aso} was performed at ambient pressure, 
and 
the true propagating vector under pressures is unknown
while the superconductivity occurs at high pressures,
we also examine the pairing symmetry with another set of parameters 
$(t_1,t_2,t_3,n)=(1.0,0.405,0.3,1.057)$ which give
 maximum values of $\chi^0(q)$  at $\mbox{\boldmath $Q$}^{\prime}
\sim (0.35\pi,0.35\pi,0.5\pi)$
as shown in Fig. \ref{fig:chi_fs}.
%%%%fig chi_fs%%%%%%%%%%%%%
%%%%%%%%%%%%%%%%%%%%%%
This ordering vector lifts the 
degeneracy of $\mbox{\boldmath $Q$}_{1}
\sim (0.5\pi,0,0.5\pi)$ and 
$\mbox{\boldmath $Q$}_{2}
\sim (0,0.5\pi,0.5\pi)$ in the $xy$ components.
 We show $\lambda$ calculated for
five irreducible representations at $\alpha =0.1$ 
in Fig. \ref{fig:lambda_fs}.
%%%%fig lambda_fs%%%%%%%%%%%%%
%%%%%%%%%%%%%%%%%%%%%%
In this case,
we again find that the A$_1$-symmetric pairing state is most likely to appear, 
which asserts the robustness of the stability of the pairing
state with the A$_1$ symmetry
against a slight change of the Fermi surface. 
B$_1$ symmetry is the second probable candidate because
the gap function $d_0^{{\rm B}_1}=\cos(2k_x)-\cos(2k_y)$
is favorable with
the propagating vector $Q^{\prime}_x=Q^{\prime}_y\sim
0.35\pi$ which could lead to the sign change
$d_0^{{\rm B}_1}(\mbox{\boldmath $k$})\cdot
d_0^{{\rm B}_1}(\mbox{\boldmath $k$}+\mbox{\boldmath $Q$}
^{\prime})<0$
on some regions $\{\mbox{\boldmath $k$}\in {\rm FS}\}$.
We calculate $\{\lambda^{\Gamma}\}$ 
for other sets of parameters with which 
the propagating vector is of the form $\mbox{\boldmath $Q$}
=(\delta,\delta,\pi/2)$ or $(\delta,0,\pi/2)$ and 
confirmed that A$_1$-symmetric superconductivity is
the most probable.
This is because $d_0^{{\rm A}_1}=\cos 2k_z$ 
depends only on $k_z$ and the changes in the $x$, $y$
components of $\mbox{\boldmath $Q$}$ do not affect 
the main scattering processes for the 
A$_1$-symmetric superconductivity.
Thus, unless the effects of pressures are not restricted
to the fluctuations in the $xy$ components of 
$\mbox{\boldmath $Q$}$, it is most stable.

\begin{figure}[tb]
\begin{center}
\includegraphics[width=0.6\linewidth]{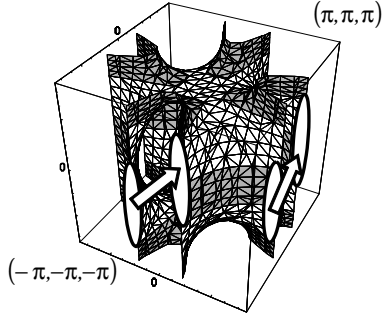}
\end{center}
\caption{Fermi surface for $(t_1,t_2,t_3,n,\alpha )=
(1.0,0.475,0.3,1.05,0)$. The gray
regions and white regions correspond to 
$d_0^{{\rm A}_1}=\cos 2k_z>0$, $d_0^{{\rm A}_1}<0$ respectively.
The line nodes of the A$_1$ gap function are located on
the boundaries between the two regions.
The white arrows connecting 
the areas enclosed by contours represent
the scattering processes associated with the propagating 
vector $\mbox{\boldmath $Q$}_{1,2+}$.}
\label{fig:node}
\end{figure}

Let us consider a possible explanation for the stability 
of the superconductivity with this symmetry in the 
case of $(t_1,t_2,t_3,n,\alpha )=
(1.0,0.475,0.3,1.05,0)$. 
We think that, in our model, taking into account only the scattering 
processes with $\mbox{\boldmath $k$}\pm \mbox{\boldmath $k$}^{\prime}
=\mbox{\boldmath $Q$}_{1,2\pm}$ gives us 
intuitive but restricted information,
because the peak 
structures of $\hat{\chi}_0(q)$ are not so sharp and 
the shape of the FS is complicated in the 3D momentum space.
Nevertheless we try to figure out how these
scattering processes on the FS contribute to the realization
of the superconductivity with the A$_1$ symmetry.
Figure \ref{fig:node} shows the FS and the signs 
of the singlet A$_1$ gap function $d_0^{{\rm A}_1}=\cos 2k_z$;
the gray and the white regions correspond to 
$d_0^{{\rm A}_1}=\cos 2k_z>0$ and $d_0^{{\rm A}_1}<0$, respectively.
The FS has a cylinder-like shape
along $z$-axis 
and there exist wide ranges of 
hot spots.
%%%%figure:node%%%%%%%%%%%%%%%%%
%%%%%%%%%%%%%%%%%%%%%%%%%%%%%%
Among these regions, the subsets of the FS connected 
to each other via the momentum $\mbox{\boldmath $Q$}$
play an important role for the superconductivity, when 
$d_0^{{\rm A}_1}(\mbox{\boldmath $k$})\cdot d_0^{{\rm A}_1}
(\mbox{\boldmath $k$}\pm\mbox{\boldmath $Q$})$ is
negative and large. Such spots might be on the sides 
of the cylinder-like FS as shown in Fig. \ref{fig:node}
(enclosed by contours) and the area of the spots 
could be large. 
Thus, the scattering processes drawn with 
white arrows in Fig. \ref{fig:node} could mediate 
the superconductivity.

%%%%%%%%%%%%%%%%%%%%%%%%%%%%%%
%related properties@@@
%%%%%%%%%%%%%%%%%%%%%%%%%%%%%%
\section{Density of states and NMR relaxation rate}
\begin{figure}[tb]
\begin{center}
\includegraphics[width=0.6\linewidth]{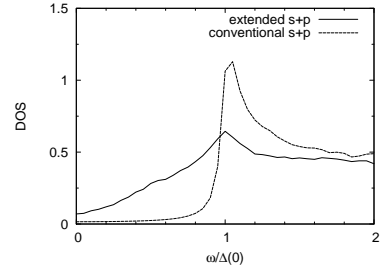}
\end{center}
\caption{The density of states in the superconducting state with
$\Delta_s(0)=2.0T_c$,
$\Delta_t(0)/\Delta_s(0)=0.01$ and $\gamma=0.01\Delta_s(0)$.
The solid line is for the A$_1$-symmetric superconducting state and
the dashed line is for the conventional $s+p$ wave state.}
\label{fig:DOS}
\end{figure}

 We now turn to the properties in 
the superconducting state: the density of states 
$\rho (\omega)$
and the NMR relaxation rate $1/T_1T$
with A$_1$ symmetry $d^{{\rm A}_1}_{\mu}$.
They are expressed as\cite{pap:Fujimoto}, 
\begin{eqnarray}
\rho(\omega)&=&\frac{1}{\pi}N_n(\omega),\\
\frac{1}{T_1T}&\propto &\int \frac{d\omega}{2\pi}
 \frac{1}{2T\cosh ^2\frac{\omega}{2T}}\left(
 \left| N_n(\omega)\right|^2
 +\left| N_a(\omega)\right|^2\right), 
\end{eqnarray}
where
\begin{eqnarray*}
 N_n(\omega)&=&-\sum_{k \tau}
 {\rm Im}G_{\tau}^{0R}\left( \omega+i\gamma,\mbox{\boldmath $k$}\right),\\
 N_a(\omega)&=&-\sum_{k \tau}
 {\rm Im}F_{\tau}^{0R}\left( \omega+i\gamma,\mbox{\boldmath $k$}\right).
\end{eqnarray*} 
The normal and anomalous Green's functions in the superconducting 
state are given by
\begin{eqnarray*}
 G^{0}_{\tau}\left( k\right)&=&\frac{i\omega_n+\xi_{k\tau}}
 {\left(i\omega_n\right)^2
 -E_{k\tau}^2},\\
 F^{0}_{\tau}\left( k\right)&=&\frac{\Delta_{k\tau}}
 {\left(i\omega_n\right)^2
 -E_{k\tau}^2},
\end{eqnarray*}
with
\begin{eqnarray*}
 E_{k\tau}&=&\sqrt{\xi_{k\tau}^2+\Delta_{k\tau}^2},\\
 \Delta_{k\tau}&=&\Delta_s\left( T\right)d_0\left( 
 \mbox{\boldmath $k$}\right)+\tau \Delta_t\left( T\right) 
 \| \mbox{\boldmath $d$}\left( \mbox{\boldmath $k$}\right)\|.
\end{eqnarray*}
We assume the $T$ dependence of the order parameters as 
$\Delta_{\mu}(T)=\Delta_{\mu}(0)\tanh \left( 
1.74\sqrt{T_c/T-1}\right)$, regarding $\Delta_{\mu}(0)$ 
and the quasiparticle damping factor $\gamma$ as fitting parameters. 
In Fig. \ref{fig:DOS}, the density of states for the extended 
$s+p$ wave superconducting state is shown, which is compared with 
that of the conventional $s+p$ wave state.
%%%%fig DOS%%%%%%%%%%%%%
%%%%%%%%%%%%%%%%%%%%%%
In contrast to the conventional $s+p$-wave state, 
the peak of $\rho(\omega)$ 
at $\omega \sim \Delta$ is largely suppressed 
and the behavior at $\omega \sim 0$ is proportional
to $\omega$ in our system. The latter aspect directly follows from 
the existence of line nodes for the A$_1$-symmetric gap
function. As for the former effect, one notices that
the strong suppression  of $\rho(\omega)$ makes its
profile quite similar to that in usual $d$-wave states.
This means that, in CeRhSi$_3$ and CeIrSi$_3$,
bulk properties which reflect 
the nodal structure of the gap function may
show no essential difference from those for $d$-wave superconductors.

\begin{figure}[tb]
\begin{center}
\includegraphics[width=0.6\linewidth]{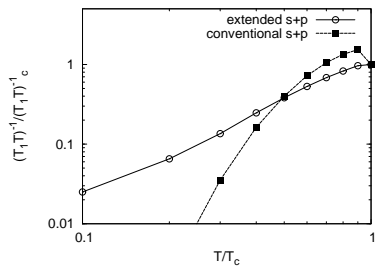}
\end{center}
\caption{NMR relaxation rate $1/T_1T$ as a function of $T$.
The amplitude of 
 the gap function at $T=0$ and the quasiparticle damping 
 factor are chosen as $\Delta_s(0)=2.0T_c$,
 $\Delta_t(0)/\Delta_s(0)=0.01$ and $\gamma=0.01\Delta_s(0).$
The line with open circles is for the A$_1$-symmetric superconducting 
state and the line with black squares is for 
the conventional $s+p$ wave state.}
\label{fig:T1}
\end{figure}

Next, we proceed to discuss the NMR relaxation 
rate $1/(T_1T)$.
Since  $1/(T_1T)$ generally depends on $N_{n}(\omega)$ and 
$N_{a}(\omega)$ 
not only at $\omega \sim 0$ but also at $\omega \sim \Delta(0)$,
the existence of line nodes does not necessarily mean that 
$1/T_1T$ in the extended $s+p$ wave superconducting state is similar to 
that in $d$-wave states.
Figure \ref{fig:T1} shows the temperature dependence of $1/(T_1T)$
normalized by $1/(T_1T)_c$ for the A$_1$-symmetric
superconducting state and the conventional $s+p$ wave state. 
By repeating similar 
calculations for several choices of
$\left(\Delta(0),\gamma \right)$,
we find that 
$1/T_1T$ with A$_1$-symmetric gap function 
has no coherence peak and behaves as $1/(T_1T) \sim T^2$, 
which is characteristic of line-node superconductors. 
%%%%figure:T1%%%%%%%%%%%%%%%%%%%%
%%%%%%%%%%%%%%%%%%%%%%%%%%%%%%%%
This behavior of the NMR $1/(T_1T)$ 
is usually typical for dominant $d$-wave superconductivity 
but not for dominant extended $s$-wave one. 
In the present system, however, the FS is highly anisotropic
and the A$_1$ gap function with line nodes 
could effectively behave like a $d$-wave gap function
on the FS.

According to the recent NMR experiments for CeIrSi$_3$\cite{com:Mukuda},
 $1/(T_1T)$ exhibits line-node behavior with 
no coherence peak, which seems not contradictory to 
our results.

%
%%%%%%%%%%%%%%%%%%%%%%%%%%%%%%%%%%%%%%%%%%%%%
%conclusion@@@
%%%%%%%%%%%%%%%%%%%%%%%%%%%%%%%%%%%%%%%%%%%%%%%%
\section{Summary}
 We have studied the pairing symmetry and 
the nature of the gap function in the superconducting state in 
the noncentrosymmetric heavy fermion superconductors 
CeRhSi$_3$ and CeIrSi$_3$. 
Solving the Eliashberg equation within RPA,
we have found that AF fluctuations could mediate the 
superconductivity with the parity mixing of the extended 
$s$-wave and $p$-wave states rather than 
the $d+f$ wave state through the Rashba spin-orbit 
interaction. 
We have confirmed that extended $s+p$ wave state is
robust against a slight change of the Fermi surface 
under pressure.
In the superconducting state,
the density of states $\rho(\omega)$ is very 
similar to that in $d$-wave superconducting states;
suppressed $\rho(\omega \sim \Delta)$ and 
$\rho(\omega \sim 0)\propto \omega$.
Furthermore, the NMR relaxation rate exhibits $1/(T_1T)\propto T^2$ 
with no coherence peak at $T_c$ as 
in the case of the usual $d$-wave superconductivity. 
Our results suggest a possible understanding of
the recent NMR experiment within the extended $s+p$ wave state.

In the present paper, the effects of the normal self-energy are not taken into
account. According to the recent experimental observations, 
the strong-coupling effect may be important in CeRhSi$_3$ and CeIrSi$_3$.
\cite{pap:Kimura3,pap:Tateiwa}
We would like to address this issue in the near future.

%%%%%%%%%%%%%%%%%%%%%%%%%%%%%%%%%%%%%%%%%%%%%%%
%%%%Acknowledgement
%%%%%%%%%%%%%%%%%%%%%%%%%%%%%%%%%%%%%%%%%%%%%%%%
\section*{Acknowledgement}
We thank M. Sigrist, N. Kimura, H. Mukuda, H. Harima,
T. Terashima, H. Yamagami, and Y. Onuki for  valuable discussions.
Numerical calculations were partially carried out at the 
Yukawa Institute Computer Facility.

%%%%%%%%%%%%%%%%%%%%%%%%%%%%%%%%%%%%%%%%%%%%%%%%%%%%
%%%%%%%%%%%%%%%%%%%%%%%%%%%%%%%%%%%%%%%%%%%%%%%%%%%%

\end{document}